\documentclass[11pt,a4paper]{article}
\usepackage[hyperref]{acl2020}
\usepackage{times}
\usepackage{amsfonts}
\usepackage{amsmath}
\usepackage{latexsym}
\usepackage{graphicx}
\usepackage{subcaption}
\usepackage{xcolor}

\usepackage{microtype}

\aclfinalcopy

\title{Tutela: An Open-Source Tool for Assessing User-Privacy \\ on Ethereum and Tornado Cash}

\author{
  \small{Mike Wu, Will McTighe, Kaili Wang}\\
  \small{Stanford University} \\ \And
  \small{Istv\'{a}n A. Seres} \\
  \small{E\"{o}tv\"{o}s Lor\'{a}nd University} \\ \AND
  \small{Nick Bax} \\
  \small{Convex Research} \\ \And
  \small{Manuel Puebla, Mariano Mendez}\\
  \small{Universidad de Buenos Aires}\\
  \small{Facultad de Ingenieria}\\ \AND
  \small{Federico Carrone, Tomás De Mattey,  Herman O. Demaestri, Mariano Nicolini, Pedro Fontana} \\
  \small{LambdaClass} \\}
\date{}

\begin{document}
\maketitle
\begin{abstract}
A common misconception among blockchain users is that pseudonymity guarantees privacy. The reality is almost the opposite. Every transaction one makes is recorded on a public ledger and reveals information about one's identity.
Mixers, such as Tornado Cash, were developed to preserve privacy through ``mixing'' transactions with those of others in an anonymity pool, making it harder to link deposits and withdrawals from the pool. Unfortunately, it is still possible to reveal information about those in the anonymity pool if users are not careful.
We introduce Tutela, an application built on expert heuristics to report the true anonymity of an Ethereum address.
In particular, Tutela has three functionalities: first, it clusters together Ethereum addresses based on interaction history such that for an Ethereum address, we can identify other addresses likely owned by the same entity; second, it shows Ethereum users their potentially compromised transactions; third, Tutela computes the true size of the anonymity pool of each Tornado Cash mixer by excluding potentially compromised transactions. A public implementation of Tutela can be found at \url{https://github.com/TutelaLabs/tutela-app}. To use Tutela, visit \url{https://www.tutela.xyz}\footnote{For related questions and inquiries, please contact the corresponding author at \texttt{wmctighe@stanford.edu}.}.
\end{abstract}

\section{Introduction}
\label{sec:introduction}

On any public blockchain, the cost of creating a new wallet is virtually zero, enabling the same entity to manage several pseudonymous addresses. The pseudonymity underpinning blockchains like Bitcoin \citep{nakamoto2008bitcoin} and Ethereum \citep{buterin2013ethereum} breeds a sense of privacy. This often leads to misuse \citep{christin2013traveling}, such as money laundering through a large number of addresses \citep{moser2013inquiry}, or unfair voting power distributed among multiple addresses owned by the same user. Thus, it is of interest in many investigations to identify addresses linked to the same entity. This is predominantly done through heuristics. Every transaction an address makes on a blockchain is recorded and public, revealing information about the underlying entity. As such, with graph analysis tools, one can cluster addresses together that, with reasonable confidence, possess the same owner.

Such anonymity tools have been widely explored for Bitcoin \cite{haslhofer2016bitcoin}, leveraging heuristics targeting the unspent transaction output (UTXO) model. However, this has limited application to more recent blockchain implementations like Ethereum, which forgo the UTXO model for an account (or sometimes balance) model.
Ethereum, in particular, has an account-based protocol that implicitly encourages an entity to reuse a handful of addresses.
As such, this poses greater challenges to user privacy than UTXO-based blockchains.

In response to this shortcoming, several coin mixing protocols have been proposed like M\"{o}bius \citep{meiklejohn2018mobius}, MixEth \citep{seres2019mixeth}, and Tornado Cash \citep{pertsev2019tornado} to obfuscate transaction tracing, the final of which is deployed in practice.
Still, new heuristics have surfaced \citep{victor2020address,beres2021blockchain} that deanonymize Ethereum users. These heuristics largely exist in academic silos, and not been combined nor demonstrated in public application.

\paragraph{Our contributions.} We develop a web application that combines several state-of-the-art heuristics to measure the anonymity of Ethereum addresses. 
To the best of our knowledge, this is the first instance to deploy these algorithms at scale.
In doing so, we create a rich depiction of user behavior and privacy.
We also propose a set of new heuristics targeted at Tornado Cash, highlighting that careless user behavior, despite using a mixer, can still reveal identity. A Python implementation is open sourced at \url{https://github.com/TutelaLabs/tutela-app} and the tool is available at \url{https://www.tutela.xyz}.

\paragraph{Paper organization.} The rest of this paper is organized as follows. We provide some pertinent preliminaries in Section~\ref{sec:preliminaries}. In Section~\ref{sec:tutela}, we provide an overview of Tutela, our developed anonymity tool. In Section~\ref{sec:data}, we describe our data processing methods and the used datasets. In Section~\ref{sec:eth}, we describe two heuristics that allowed us to cluster Ethereum addresses that are likely owned by the same entity. In Section~\ref{sec:tornado}, we assess the privacy guarantees of Tornado Cash applying five novel heuristics. In Section~\ref{sec:analysis}, we provide a quantitative analysis. Finally, we conclude our work with some discussions and future work in Section~\ref{sec:discussion}.
\section{Preliminaries}
\label{sec:preliminaries}
In this section, we provide some background on Ethereum and its account model. For a deeper dive into Ethereum, we refer the reader to~\cite{antonopoulos2018mastering}. Moreover, we briefly describe the high-level workings of Tornado Cash.

\subsection{Ethereum and the Account Model}
Ethereum is the second-largest cryptocurrency platform in terms of market capitalization.\footnote{For more details, see: \url{https://coinmarketcap.com/currencies/ethereum}.} However, it is the largest smart contract platform, and it is also the most used public blockchain for settling transactions. Therefore, it is imperative to understand better and assess the privacy guarantees of Ethereum quantitatively.

Ethereum employs the \textit{account model}. There are two types of accounts: externally owned accounts (EOA) and contract accounts. EOAs are controlled by users via a cryptographic key-pair (a private and a public key) owned by them. The private key of the EOA enables users to send transactions from that account, while the public key is used to derive an address for the EOA (the hash of the public key is its address). On the other hand, contract accounts are controlled by contract code. The contract account's address is the hash of their contract code. Accounts are referred to by their addresses (a pseudonym). Contract accounts cannot initiate transactions. However, EOAs can send transactions to contract accounts that can run the code of the contract account. In the following, we will use the words accounts and addresses interchangeably. 


Ethereum's account model has several implications from a privacy point of view. First, the account model incentivizes the reuse of accounts across many transactions. Imagine the following scenario. Alice owns an EOA with address A and a balance of three ether. Alice wishes to buy a product from Bob for two ether. Alice sends a transaction worth two ether to Bob. After this purchase, address A has a balance of one ether. Therefore, if Alice wants to spend her change, she necessarily needs to send a transaction \emph{again} from her address A. Address reuse facilitates the profiling of rich transaction histories: complete financial history,  list of all counterparties, time-of-day activity, and more. Furthermore, account reuse implies that most users only own a handful of accounts. Finally, the accounts owned by the same user can effectively be clustered; see Section~\ref{sec:eth}. 

\subsection{Preserving Privacy on Ethereum: Tornado Cash}
Users can break links with their transaction history using a so-called mixer contract. Tornado Cash (TC) is the most widely used, non-custodial mixer on Ethereum. TC works as follows. Users deposit equal amounts to a TC smart contract (e.g. 1 ETH as shown below). After some time, users can withdraw their funds from the mixer contract to \emph{a freshly generated EOA} by providing a zero-knowledge proof that proves that the withdrawing user is one of the depositors. Therefore, the withdrawing EOA has enhanced its privacy since it has become unlinkable to any unique depositor EOA. Note that each TC contract applies a fixed denomination for the mixed funds (e.g. 1 ETH). Otherwise, linking deposits to withdraws would be trivial.

\begin{figure}[h!]
\includegraphics[width=\linewidth]{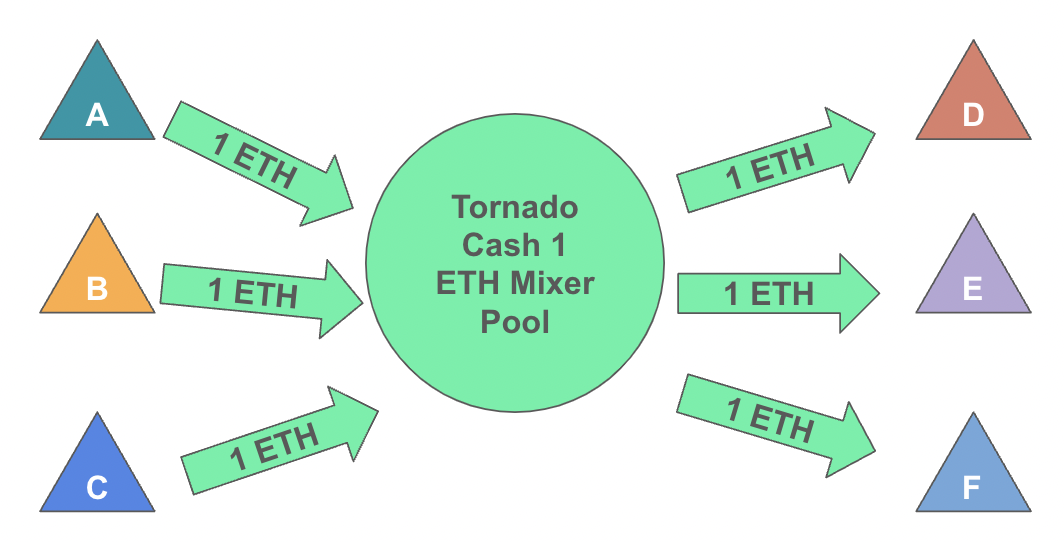}
\caption{Example of the Tornado Cash 1 ETH pool: addresses A through F deposit to and withdraw from the pool. It quickly becomes impossible to associate withdraw and deposit transactions given a growing mixer.}
\label{fig:demo}
\end{figure}

A user's anonymity is defined by the number of equal user deposits in a given pool. This is the pool's \emph{Anonymity Set}. In the example above, D’s withdrawal could have come from A, B or C, so the anonymity set is 3 and the probability of correctly guessing the deposit / withdrawal connection is 1/3.

The more users that deposit in the pool, the greater the number of people that a withdrawal could have come from. If you add a fourth deposit to the example above, the probability of being correctly detected decreases to 1/4.
However, there are lots of ways users can compromise their privacy. If you can link A’s deposit to E’s withdrawal, then the pool’s anonymity set decreases from 3 to 2. This means the probability of correctly guessing your deposit / withdrawal connection increases to 1/2 because any withdrawal could only have come from B or C’s deposits.

Tornado Cash provides mixing contracts (otherwise known as mixing pools) for Ether and several Ethereum-based tokens (e.g., DAI, USDC, wBTC, etc.) in different denominations (1, 10, 100, etc.).

\section{Tutela Overview}
\label{sec:tutela}

Tutela\footnote{At the time of publication, Tutela is hosted at \url{https://www.tutela.xyz}.}, latin for protection, is a web application that has three functions, it inform users which of their Ethereum addresses are affiliated, how they are linked and audits the anonymity sets of Tornado Cash Pools. More on each of these below.

\subsection {Ethereum Address Clustering}

Users can search an ethereum address (e.g. 0x...) or ENS (e.g. tutela.eth), and receive a summary of its anonymity. 

\begin{figure}[h!]
\includegraphics[width=\linewidth]{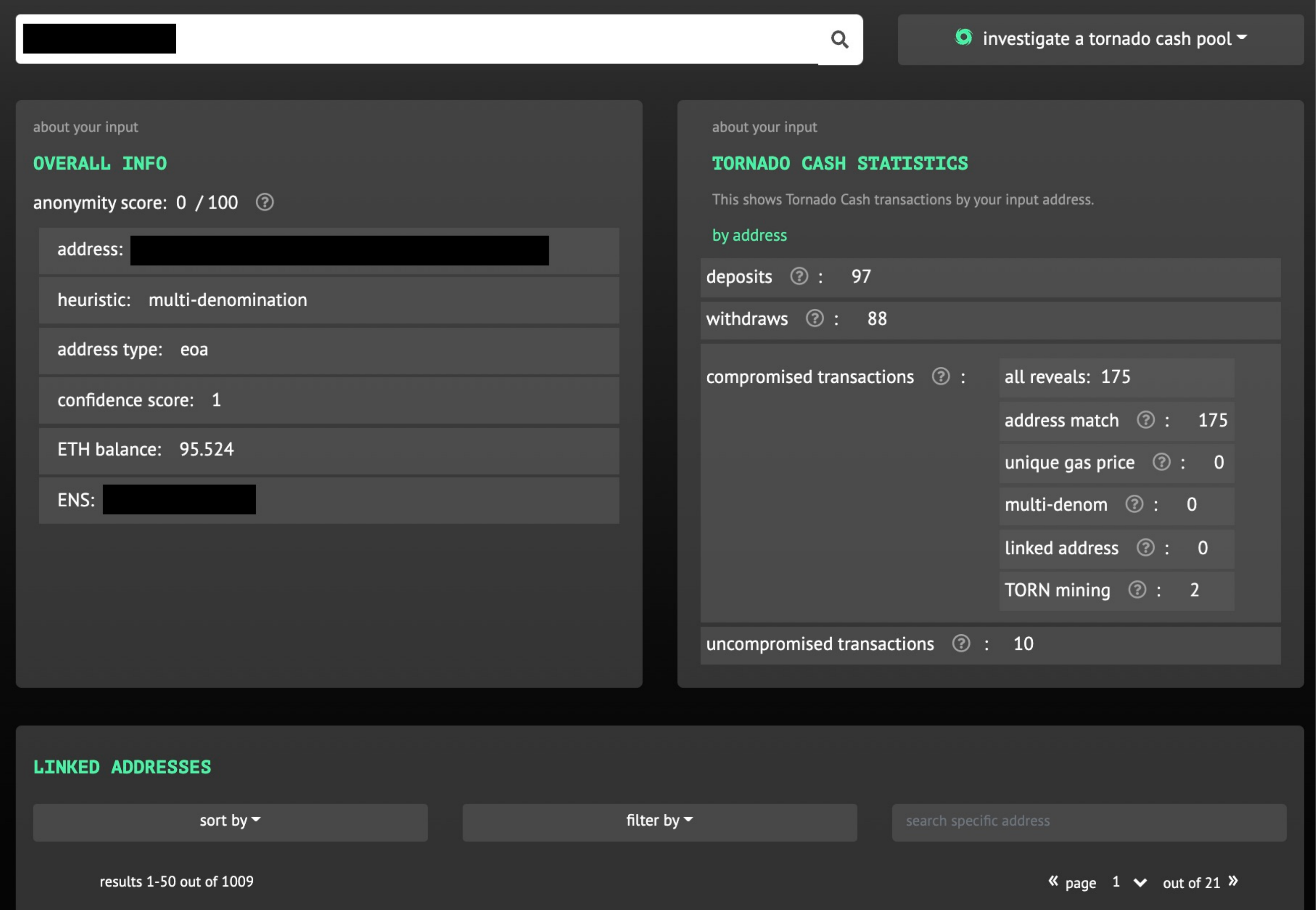}
\caption{Tutela address page when searching an Ethereum address. This address is also a TC user.}
\end{figure}

We summarize the main functionalities shown in Figure~\ref{fig:demo}. The response page is separated into three sections. The top left section summarizes the anonymity of the searched address. It contains an ``anonymity score'' out of 100, where a lower number represents less anonymity. In this example, the searched address has an anonymity score of 0, representing a large amount of leaked identity information. Other information, such as balance in ETH or ENS names, are shown when relevant.

The top right section is only populated if the searched address is a Tornado Cash user. In this example, the searched address has deposited 97 times to a Tornado Cash pool and withdrawn 88 times. Interestingly, we find that through heuristics, we are able to tie 87 of those withdraw transactions to deposit transactions, thereby meaningfully reducing the useful size of the Tornado Cash pool. See Section~\ref{sec:tornado} for more details.

The bottom section labeled ``Linked Addresses'' shows a list of addresses clustered with the searched one. Each item in this list is denoted as either an externally owned address (EOA), a deposit address, or an exchange address. Each item also contains a confidence score denoting the strength of association with the searched address, and the heuristic that bound it to the searched address.
This example shows a thousand clustered addresses representing an entity with a large wallet portfolio (e.g., bot) -- hence why the anonymity score is zero.

\subsection{Ethereum Address Reveals}

\begin{figure}[h!]
\includegraphics[width=\linewidth]{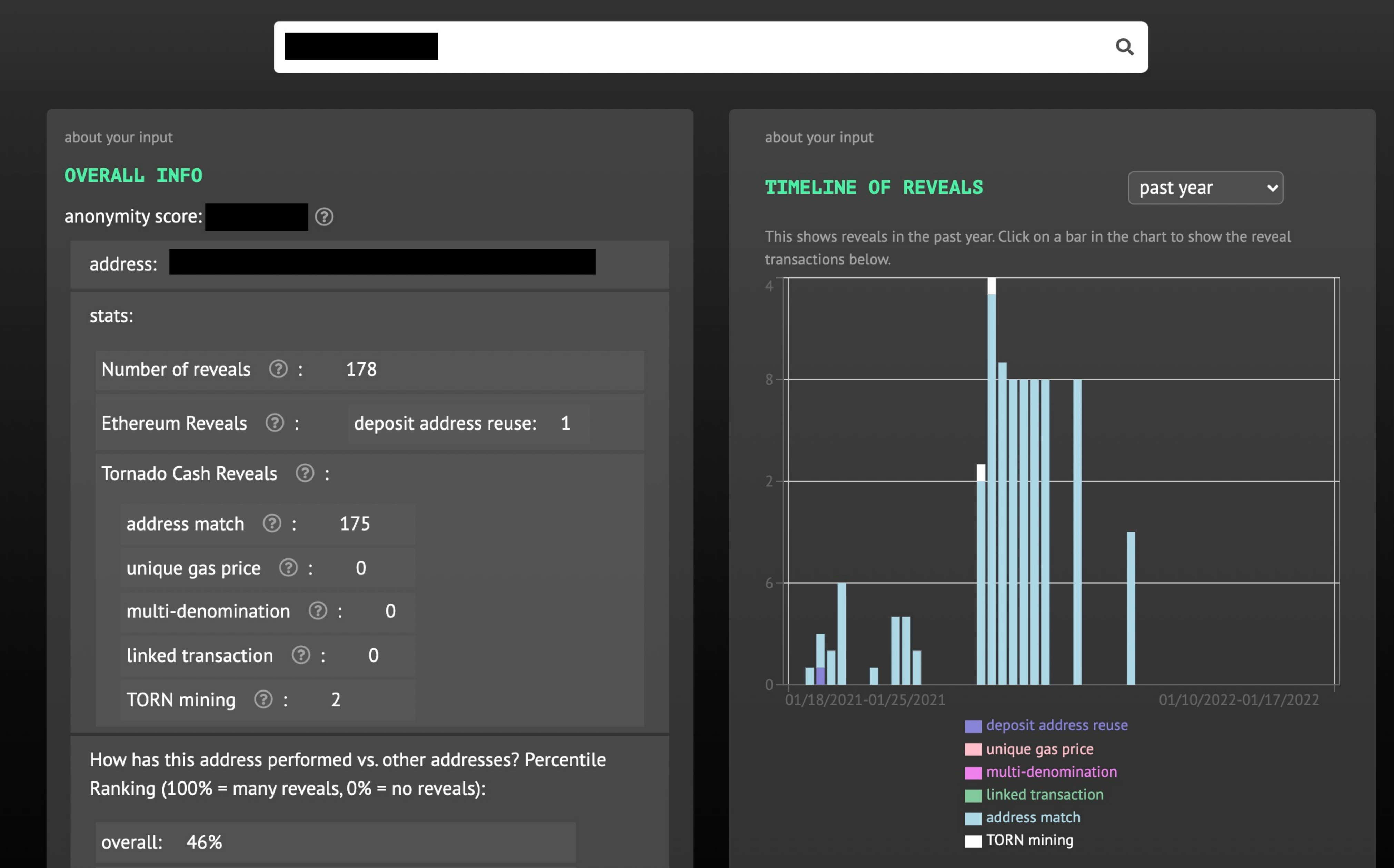}
\caption{Tutela transaction page when searching an Ethereum address. This address is also a TC user.}
\label{fig:demo3}
\end{figure}

To see a history of when an Ethereum user potentially committed reveals, users can select the transactions tab on the landing page and input an Ethereum address or ENS. The right hand side of Figure~\ref{fig:demo3} shows a graph of when these revealing transactions occurred where the x-axis denotes the weeks prior to the current date. Each bar in the graph corresponds to an individual week and upon clicking on each bar, Tutela will show the details of the potentially revealing transactions underneath. On the left hand side, users can see statistics on their potential reveals as well as a comparison to the average Tornado Cash user.

\subsection{Tornado Cash Anonymity Set Auditor}

The Tornado Cash Anonymity Set Auditor, computes the five heuristics described in Section~\ref{sec:tornado} for each Tornado Cash pool to determine how many potentially compromised deposits are in each pool.

\begin{figure}[h!]
\includegraphics[width=\linewidth]{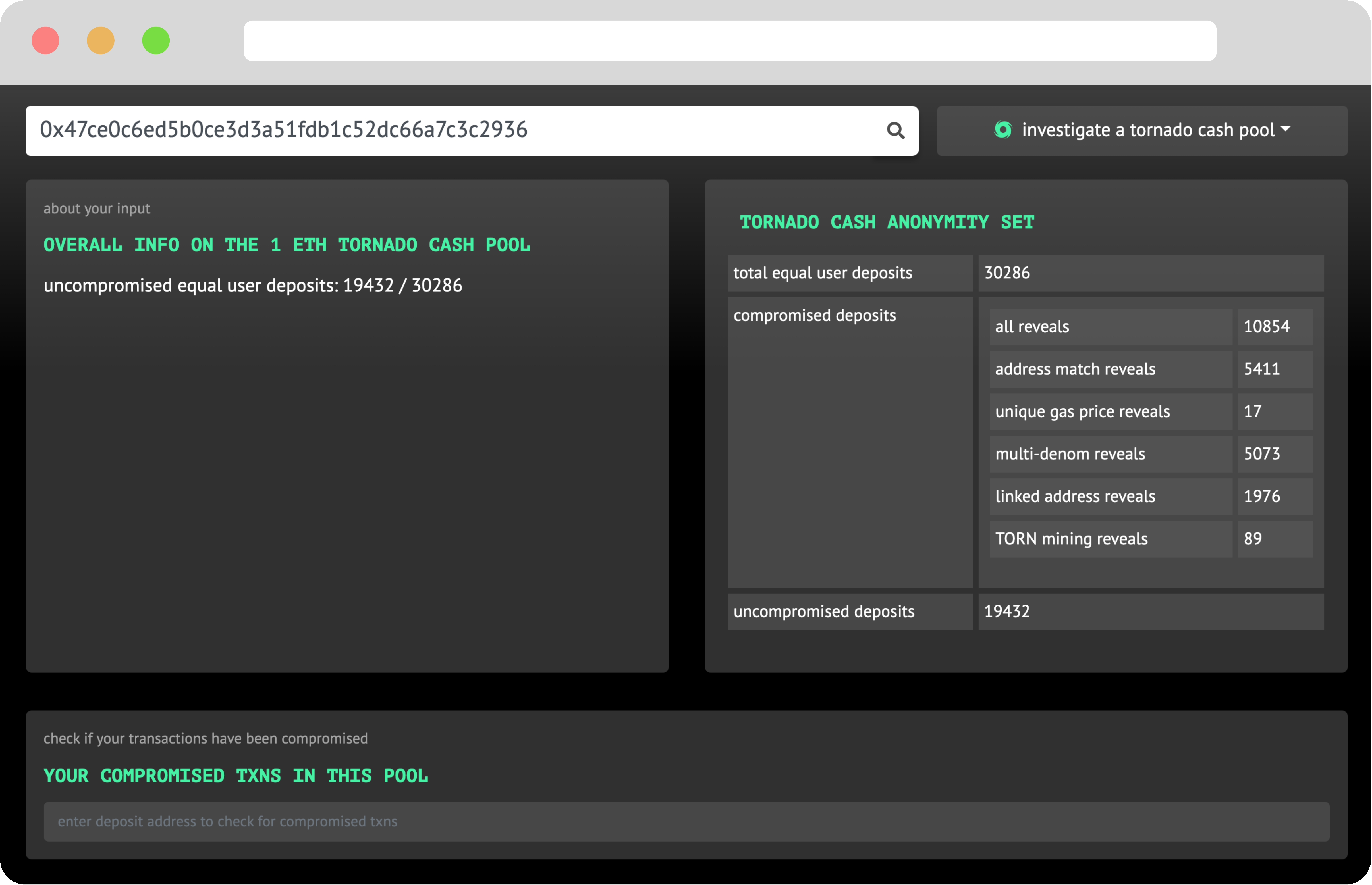}
\caption{Tutela interface when searching a Tornado Cash pool.}
\label{fig:demo2}
\end{figure}

If a user selects a Tornado Cash Pool from the dropdown list on the landing page, the top right of the results page will show the headline number of equal user deposits in a Tornado Cash pool, otherwise known as its anonymity set. Below that, we compute the ``true'' anonymity set size for the pool, subtracting out all equal user deposits that may have been compromised through our heuristics. This gives the number of “uncompromised deposits”. At the time of publication, the Tornado Cash website reports only the full anonymity set size, not taking into account potential compromises. 

At the bottom of Figure~\ref{fig:demo2}, users can supply an address to check, \textit{in a private way}, if it has made any compromising transactions, inspired by the popular website ``Have I been pwned?''\footnote{See \url{https://haveibeenpwned.com}.}.

A concrete use case of Tutela is to adjust anonymity set to protect the privacy of Tornado Cash users more faithfully. As described in Section~\ref{sec:preliminaries}, if a pool has many compromised deposits, it provides less privacy than a user believes.
\section{Data and Setup}
\label{sec:data}

We discuss the data sources used to build Tutela.

Ethereum transactions are downloaded from the \texttt{crypto\_ethereum} dataset using BigQuery, including all transactions from August 7th, 2015 to October 1st, 2021\footnote{In the web application, Tutela is updated weekly.}.  In total, this amounts to 4 terabytes of data with over 1B rows.
In addition, we assume access to a list of known addresses obtained from a public Kaggle challenge\footnote{See the list of labelled Ethereum addresses found at \url{https://www.kaggle.com/hamishhall/labelled-ethereum-addresses}.}, containing almost 20,000 labelled addresses corresponding to different centralized exchanges, decentralized exchanges, relayers, DeFi applications, and more.
This list will be used to identify exchange addresses for heuristics and apply known constraints on the inferred identity of clustered addresses.

Additionally, we create a partition of the transaction data from \texttt{crypto\_ethereum} pertaining to Tornado Cash pools. This is done by checking that the address receiving a transaction is a Tornado Cash smart contract (taken from the BigQuery dataset \texttt{tornado\_cash\_transactions}). To capture the transactions executed by the Ethereum virtual machine (e.g., through a smart contract), we use the \texttt{crypto\_ethereum.traces} table. In the special case that a withdrawal from a Tornado Cash pool is made by a relayer, we decode the input code using the contract ABI to find the recipient address. In total, we uncover around 97,365 deposit and 83,782 withdraw transactions across all pools. These two transaction sets will be used for Tornado Cash-specific heuristics.
\section{Ethereum Heuristics} 
\label{sec:eth}

Using this large dataset, we describe two Ethereum-wide heuristics used to cluster together addresses potentially belonging to the same entity.

\subsection{Deposit Address Reuse}
\label{sec:dar}

Deposit address reuse (DAR) links together EOAs through the usage of a centralized exchange (CEX). We refer to the original paper \citep{victor2020address} for a detailed description but provide an overview here.

\begin{figure}[b!]
\includegraphics[width=\linewidth]{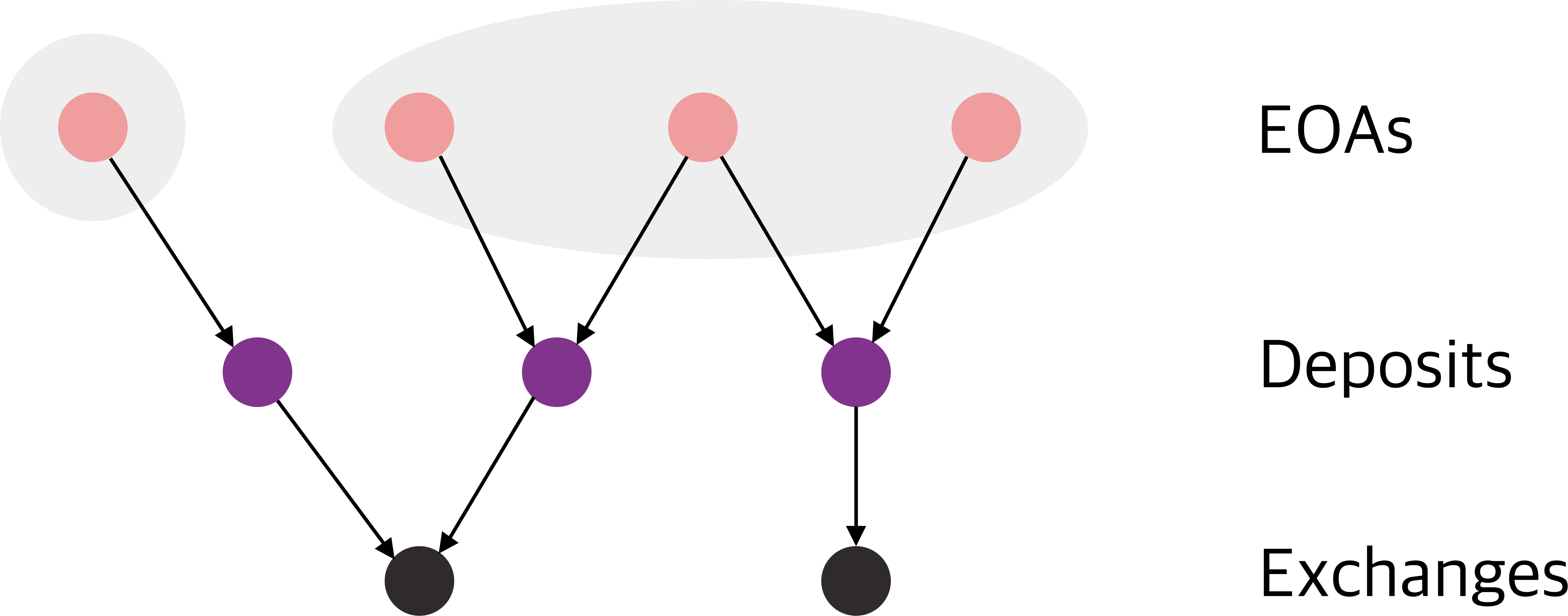}
\caption{Graph of transactions between EOA, deposit, and CEX addresses. A cluster is defined as a weakly connected component in an undirected subgraph containing only EOA and deposit nodes. The gray circles show EOA addresses in two clusters.}
\label{fig:dar}
\end{figure}

When users deposit Ether to a CEX, the exchange typically creates ``deposit addresses'', which receive the assets from the EOA and then forward these funds to the main addresses associated with the CEX. Critically, these deposit addresses are created per customer, not per address. That is, multiple addresses that send funds to the same deposit address are highly likely to be owned by the same entity. However, these deposit addresses are not known. The DAR algorithm seeks to identify deposit addresses through heuristics.
It uses two hyperparameters: the maximum amount difference $\alpha$ and the maximum time difference $\tau$ between two transactions: a ``receiving'' transaction from an EOA to a suspected deposit address, and a ``forwarding'' transaction from a deposit to a main exchange address (we assume access to a list of main CEX addresses). The intuition is that true deposits likely forward funds quickly, and differences in amount should only be due to transaction fees. We exclude known entities (e.g. CEX, DEX, miner, etc.) from being classified as deposit addresses.

Then, we can search for $($EOA, deposit, exchange$)$ tuples that match the constraints set by $\alpha$ and $\tau$.
If we construct an undirected graph with addresses as nodes and edges between associated EOA and deposits, a ``cluster'' is defined as a weakly connected component. This definition captures interesting cases where a single entity has multiple EOAs that send to different deposits of multiple CEXs (see Figure~\ref{fig:dar}). In practice, we pick $\alpha = 0.01$ and $\tau = 3,200$ \citep{victor2020address}.

For each cluster, we would like to assign a confidence score, representing our belief in this cluster representing a single entity. We note that any uncertainty in clustering must come from uncertainty in defining deposit addresses. So, for any address in a cluster $C$, we assign  confidence as the average confidence for deposits in $C$. Now, for any deposit $v$, we define confidence as:
\begin{equation*}
\kappa(v) = 1 - \left\{ \frac{1}{2}\left(\frac{|a_f - a_r|}{\alpha}\right) + \frac{1}{2}\left(\frac{t_f - t_r}{\tau}\right) \right\}
\end{equation*}
where $a_f$ and $a_r$ are the amounts from the forwarding and receiving transactions, respectively. Similarly, $t_f$ and $t_r$ are the forwarding and receiving block numbers.
A larger difference in either amount or time would decrease the confidence, which is bound to be between 0 and 1.

\subsection{Learned Node Embedding}

The benefit of DAR is interpretability: we can understand (to a degree) why the algorithm believes two addresses are linked. However, this comes at the cost of recall and the ability to identify dense clusters. DAR searches for a very specific behavior, and most addresses will not be in a cluster of greater than size 1. Initial feedback from Tutela users reported limited success in finding clusters.
To supplement DAR, we consider a second Ethereum-wide heuristic (NODE) that projects addresses to points in a low-dimensional vector space based on who it transacts with. In this vector space, addresses belonging to the same entity should be close together in Euclidean distance.

Consider constructing an undirected graph $G(V, E)$ from all Ethereum transactions, where nodes $V$ are composed of distinct addresses, and an edge is placed between two nodes if there is a transaction between them. Each edge is given a weight $w: V \times V \rightarrow \mathbb{N}$ designating the number of interactions between two addresses. For instance, and if Alice has sent Bob 1 ETH five times,  then $w(\textup{Alice}, \textup{Bob}) = 5$. Note that this graph is distinct from the one used in Section~\ref{sec:dar}.

At this abstraction layer, we seek to learn a ``node embedding function'' $f: V \rightarrow \mathbb{R}^d$ that projects a node to a $d$-dimensional vector representation. Importantly, we want this embedding to capture semantic information about the node, such as which other addresses it frequently interacts with. To do so, we leverage popular graph representation learning algorithms \citep{grover2016node2vec,rozemberczki2018fast}. In particular, we focus on Diff2Vec \citep{rozemberczki2018fast}, which has been applied to blockchain transactions \citep{beres2021blockchain}, though not at scale.

The intuition of Diff2Vec is to summarize a node by its neighborhood through a diffusion-like random process. Specifically, there are four steps to Diff2Vec: (1) generating a ``diffusion graph'', (2) sampling a ``node sequence'', (3) extracting features, (4) learning a neural network embedding. We briefly summarize each step below.

\paragraph{Step One} Fixing a starting node $v \in V$, we initialize a diffusion subgraph $\tilde{G}$ containing only $\{v\}$. Randomly sample two nodes, $u$ from $\tilde{G}$ and $w \in \textup{Neighbors}(u, G)$, where $G$ is the original graph. Add node $w$ and the edge $(u, w)$ to $\tilde{G}$. Repeat until $\tilde{G}$ has $l$ nodes, where $l$ is a hyperparameter representing the amount of information we want to capture in our eventual node embedding. A larger $l$ may capture a large neighborhood but sacrifice in granularity.

\paragraph{Step Two} Given $\tilde{G}$, we generate a sequence $s = (v_1, v_2, \ldots)$ recording the nodes visited during an Euler walk. To do this, we must ensure that $\tilde{G}$ is Eulerian, which holds if every node has an even degree. A simple way to achieve this is to double each edge in $\tilde{G}$. At this point, we can summarize a node $v \in V$ by multiple sequences $S = (s_1, s_2, \ldots)$, each from a random walk.

\paragraph{Step Three} Given a set of sequences $S$, we aim to produce a single feature vector based on frequencies of which vertices appear near each other in sequences $s \in S$. Specifically, fix a node $v$. Then, pick a window size $h$.
Count how many times other nodes appear within $h$ positions before and after when $v$ appears, summed over all sequences $s \in S$. This will result in $2h$ vectors each of length $|V|$ --- $2$ from counting before \textit{and} after; $h$ from counting frequencies 1 to $h$ positions away from $v$; $|V|$ since each vector stores counts of all nodes in $V$. Denote this feature vector by $y(v) \in \mathbb{R}^{2h|V|}$.

\paragraph{Step Four} Finally, we wish to compress the feature vector from Step Three to a $d$-dimensional continuous space. To do this, we train a two layer perceptron $f_\theta = f_1 \circ f_2$ using stochastic gradient descent \citep{goodfellow2016deep,kingma2014adam}. Define $f_1: \mathbb{R}^{|V|} \rightarrow \mathbb{R}^d$ and $f_2: \mathbb{R}^d \rightarrow \mathbb{R}^{2h|V|}$, with a ReLU nonlinearity in between. The input to the network is a one hotted representation of the current node $v$.
To train the parameters $\theta$, we optimize the objective:
\begin{equation}
  \mathcal{L}(v; \theta) = \texttt{dist}(f_\theta(\texttt{one\_hot}(v)), y(v))
  \label{eq:obj}
\end{equation}
where $\texttt{dist}$ is a distance function. Examples include Euclidean distance, or a cross entropy loss.
That is, Equation~\ref{eq:obj} optimizes $f_\theta$ to predict the frequency vector. Then, assign $f_1(v) \in \mathbb{R}^d$ to be the final embedding for $v$.\newline

\begin{figure}[h!]
  \centering
  \hfill
  \begin{subfigure}[b]{0.25\linewidth}
  \centering
  \includegraphics[width=\textwidth]{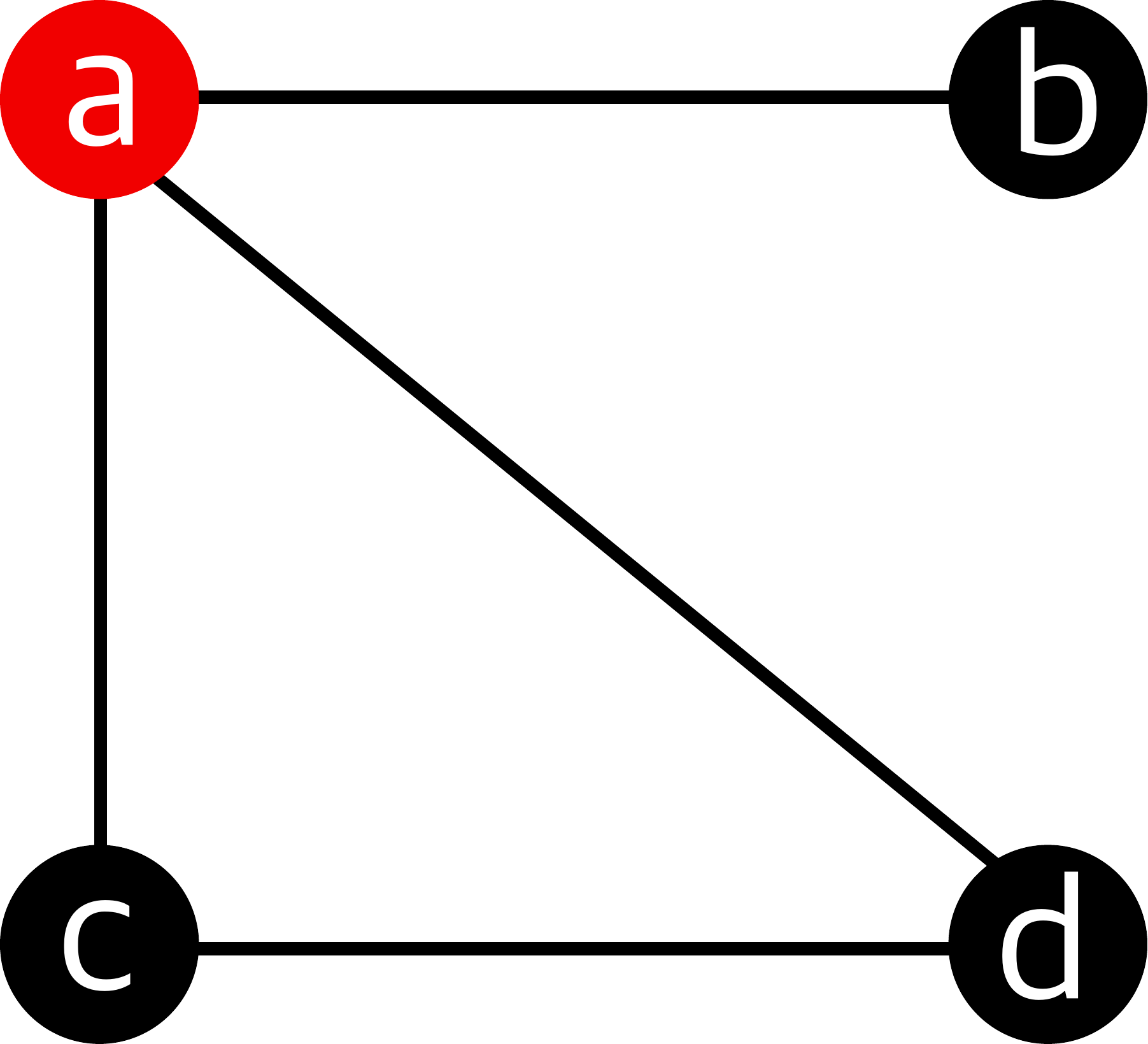}
  \caption{Subgraph $\tilde{G}$}
  \end{subfigure}
  \hfill
  \begin{subfigure}[b]{0.4\linewidth}
  \centering
  \includegraphics[width=\textwidth]{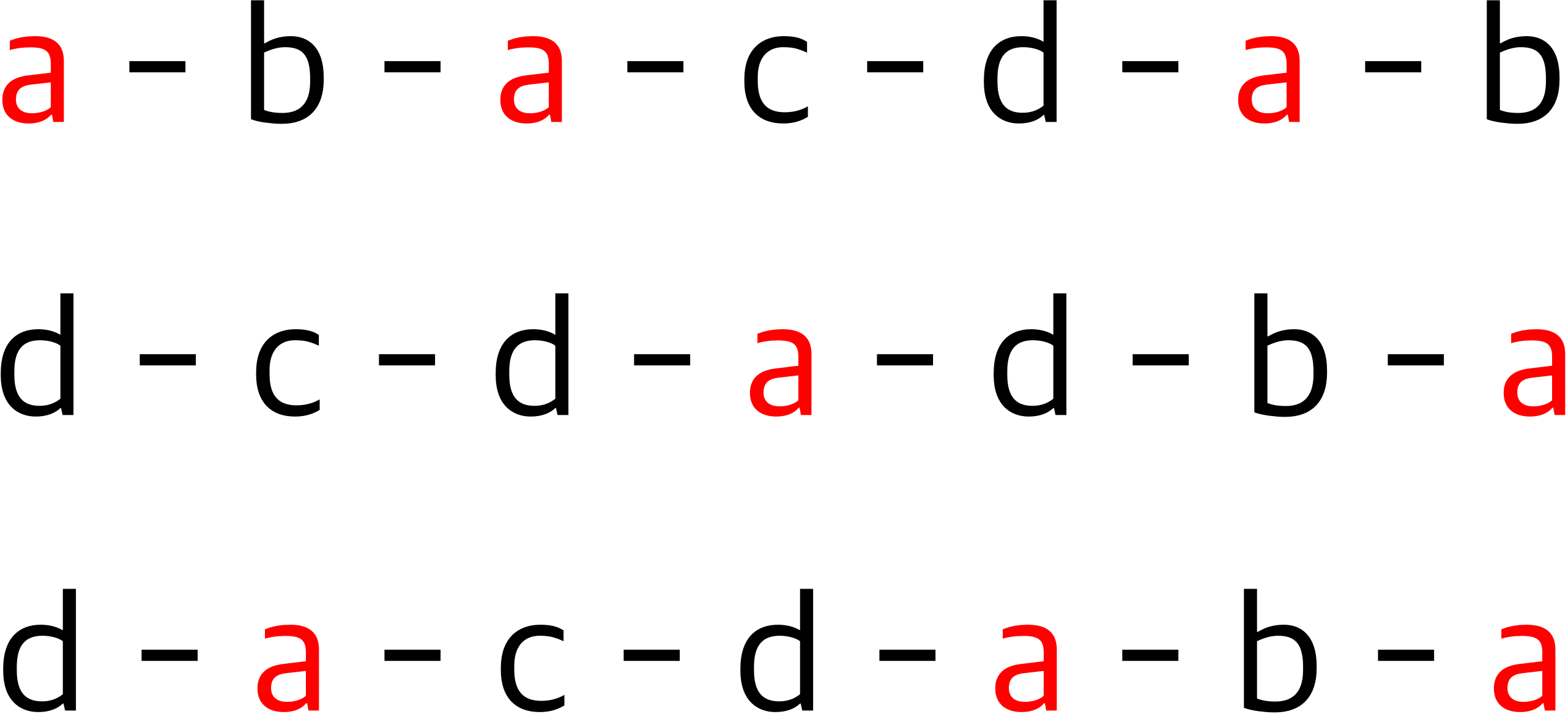}
  \caption{Sequences $S$}
  \end{subfigure}
  \hfill
  \par\bigskip
  \hfill
  \begin{subfigure}[b]{0.4\linewidth}
  \centering
  \includegraphics[width=\textwidth]{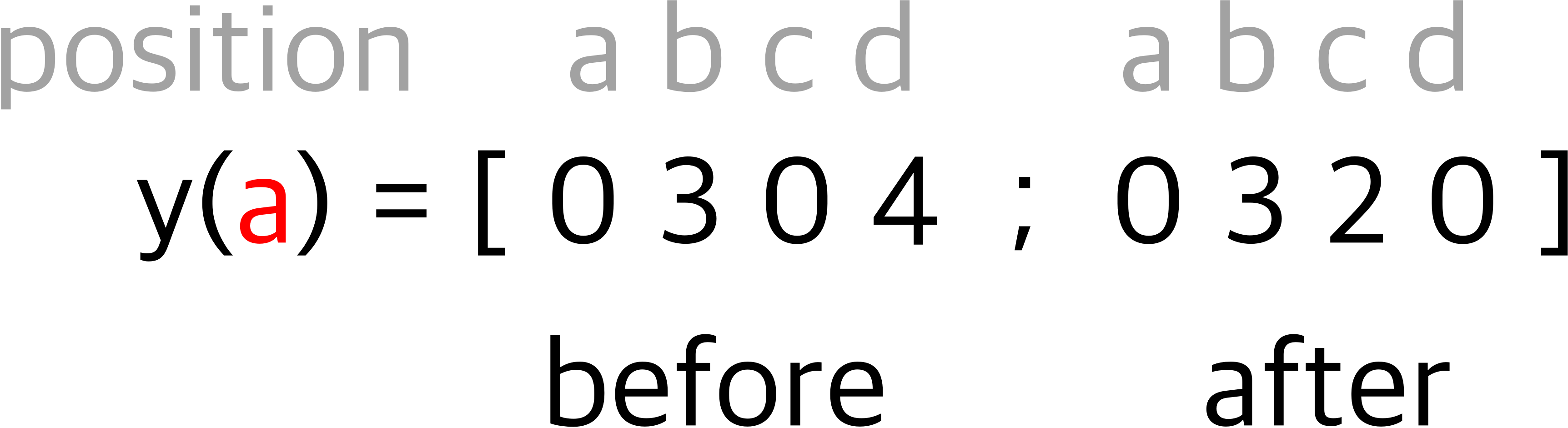}
  \caption{Feature $y(a)$}
  \end{subfigure}
  \hfill
  \begin{subfigure}[b]{0.4\linewidth}
  \centering
  \includegraphics[width=\textwidth]{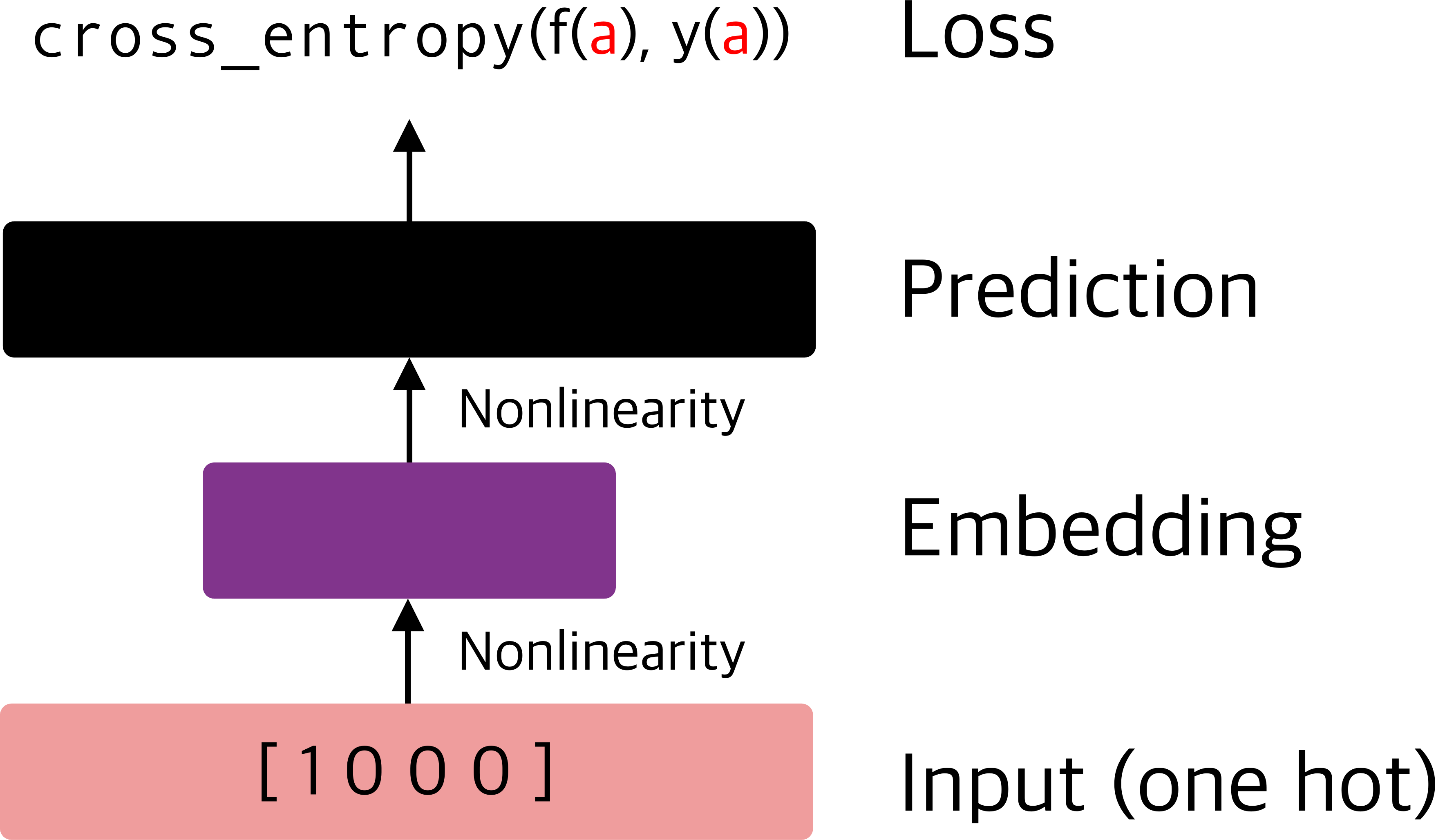}
  \caption{Word2Vec $f_\theta$}
  \end{subfigure}
  \hfill
\caption{The four steps of the Diff2Vec algorithm, focusing on the individual node $a \in V$.}
\label{fig:diff2vec}
\end{figure}

Steps three and four amount to Word2Vec\footnote{We use \texttt{gensim.models.word2vec}.} \citep{mikolov2013efficient,mikolov2013distributed}. In practice we optimize with SGD for 5 epochs with a learning rate of 0.025. We set $d = 128$, $l = 40$, and $h = 5$. Given an address $v \in V$, to find its cluster, we can search for the closest $k$ vectors in $\mathbb{R}^d$. In practice, we accomplish this efficiently using FAISS \citep{johnson2019billion}. Unlike DAR, this heuristic will always return $k$ addresses. We score the confidence of an address $u$ in the cluster by the inverse of its Euclidean distance to the embedding of $v$:
\begin{equation*}
\kappa(u) = \frac{1}{\|f_\theta(\texttt{one\_hot}(u)) - f_\theta(\texttt{one\_hot}(v))\|_2}
\end{equation*}
since a smaller distance (closer to 0) represents more semantic similarity of $u$ to $v$.

\subsection{Anonymity Score}
\label{sec:anonymityscore}

Given a query address $v$, using one or both heuristics, we obtain a set of clustered addresses $C$ with confidence scores $\kappa$ for every member. We wish to compute a statistic summarizing the anonymity of the query address, such that a larger cluster reveals more identity information, hence lower anonymity. We define the anonymity score as:
\begin{equation}
\texttt{anonymity}(v) = 1 - \texttt{tanh}(\beta * \kappa(v) * |C|)
\label{eq:anonymity}
\end{equation}
where $\beta$ is a hyperparameter controlling slope. A larger value for $\beta$ more aggressively penalizes larger clusters. We chose $\beta = 0.1$. This anonymity score is a summary statistic representing how easily connected an address is to other addresses potentially owned by the same entity.
\section{Tornado Cash Heuristics}
\label{sec:tornado}

Focusing on the subset of Ethereum transaction data involving Tornado cash deposits and withdrawals, we discuss a five heuristics for identifying compromised deposits.

\subsection{Address Match}

\begin{figure}[h!]
\centering
\includegraphics[width=0.75\linewidth]{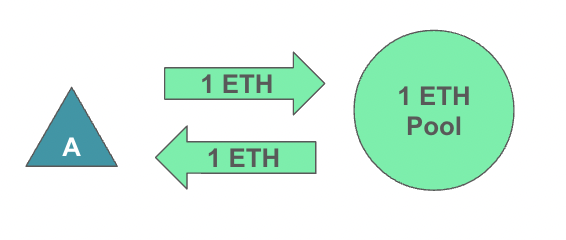}
\caption{Address match -- The triangle represents a single address withdrawing and depositing to a TC pool.}
\label{fig:tornado}
\end{figure}

Suppose the address making a deposit transaction to a Tornado Cash pool matches the address making a withdrawal transaction (from the same pool). In that case, the two transactions can be linked, and the corresponding deposit is compromised as the user identity may be revealed. These may be TORN yield farmers who deposit and withdraw to the same address and are only profit-motivated or clumsy Tornado cash users.

\subsection{Unique Gas Price}

\begin{figure}[h!]
\centering
\includegraphics[width=0.7\linewidth]{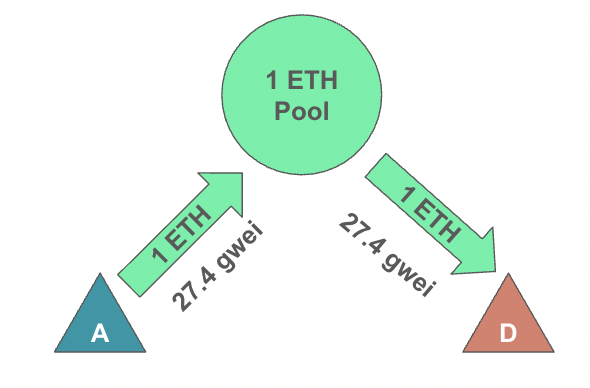}
\caption{Unique gas price -- two addresses depositing and withdrawing with the same 27.4 gwei gas price.}
\label{fig:tornado}
\end{figure}

Prior to EIP-1559, Ethereum users could specify the gas price when making a deposit or withdrawal to a Tornado Cash pool. Those who do so tend to specify gas prices that are identical for deposit and withdrawal transactions. User-specified gas prices follow common patterns such as being round numbers or being unique to an individual. Care must be taken to remove transactions made by a relayer service which may set gas prices as well. Relayer-specified prices also tend to be oddly specific and not round numbers. In practice, relayers can be filtered out by decoding the input code from Ethereum transaction data.

\subsection{Linked ETH Addresses}

\begin{figure}[h!]
\centering
\includegraphics[width=0.6\linewidth]{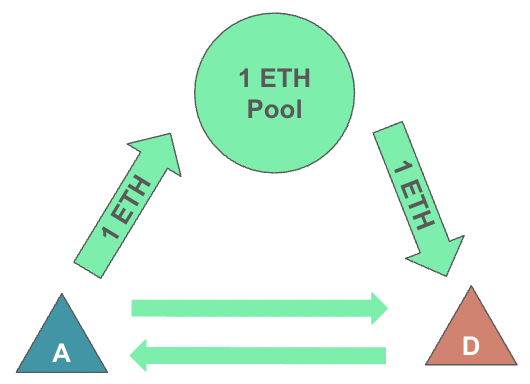}
\caption{Linked ETH addresses -- the green arrows represent interactions between two addresses A and D outside of TC. Addresses A and D deposit and withdraw from the same Tornado Cash pool, respectively.}
\label{fig:tornado}
\end{figure}

This heuristic aims to link withdraw and deposit transactions on Tornado Cash by inspecting ETH (non-Tornado Cash) interactions. This is done by constructing two sets, one corresponding to the unique Tornado Cash deposit addresses and one to the unique Tornado Cash withdraw addresses, to then make a query to reveal transactions between addresses of each set: when at least three such transactions are found, the withdraw and deposits addresses will be considered heuristically linked in Tornado Cash. The more transactions found, the more confident we are in the link.

\subsection{Multiple Denomination}

\begin{figure}[h!]
\centering
\includegraphics[width=0.6\linewidth]{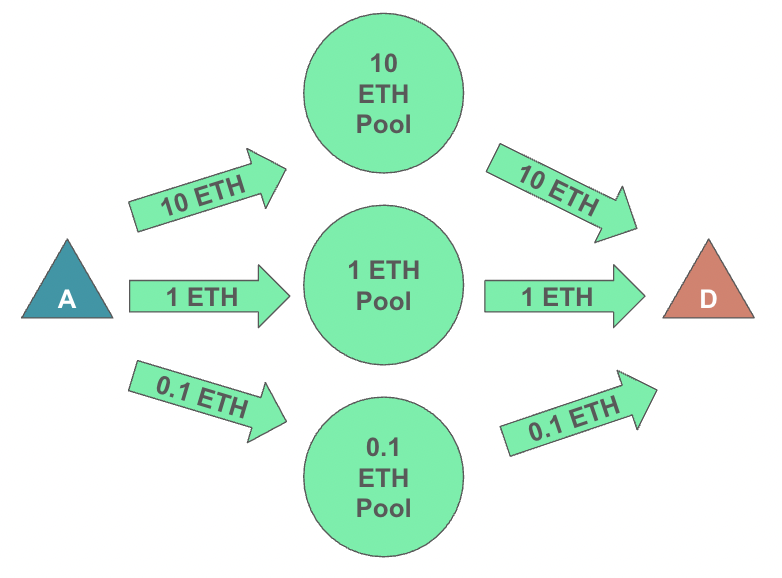}
\caption{Multi Denomination -- Addresses A and D deposit and withdraw the same number of times from the same three Tornado Cash pools, respectively.}
\label{fig:tornado}
\end{figure}

Previous heuristics examine isolated pairs of deposit and withdraw transactions. This heuristic, however, studies an address' history of transactions. We compute the portfolio of an address' withdrawals across Tornado Cash pools, hence ``multi-denomination''. For instance, Alice may have withdrawn from the 1 ETH pool twice, the 0.1 ETH pool five times, the DAI pool once. Then we search for all addresses whose portfolio of deposit transactions is \textit{exactly the same} as Alice' withdrawal portfolio. An address depositing into the 1 ETH pool three times, the 0.1 ETH pool five times, and the DAI pool once would be a valid match. (Optionally, we can relax this to search for deposits with at least as many as Alice' withdrawals, though this risks more false positives). All Tornado Cash transactions under the matched deposit and withdrawal address are linked to each other.

We ignore all addresses that make fewer than three transactions to Tornado Cash pools, and ignore all addresses that interact with only one pool. Additionally, to reduce the likelihood of false positives, we further constrain deposits within a multi-denomination reveal to occur within a 24 hour window; similarly we separately constrain withdraws within a multi-denomination reveal to occur within a 24 hour window. This intuition being that careless users will deposit or withdraw all at once.

\subsection{TORN Mining}

\begin{figure}[h!]
\centering
\includegraphics[width=0.6\linewidth]{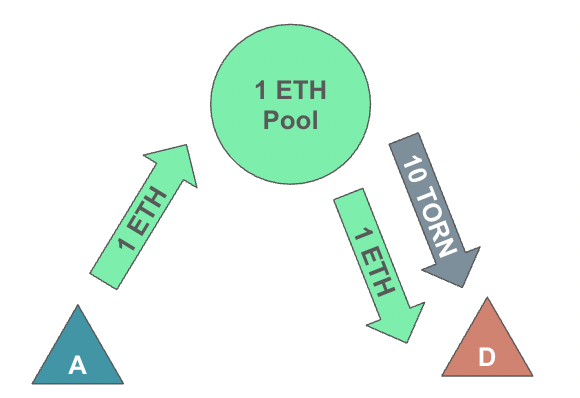}
\caption{TORN Mining -- Address D was given 10 TORN upon withdrawing from the 1 ETH pool in return for anonymity points, linking address D to a deposit 100 blocks prior. Searching  records, only address A deposited in the 1 ETH pool 100 blocks prior, compromising address D. Note that the numbers presented here are for explanatory purposes.}
\label{fig:tornado}
\end{figure}

In February 2021, Tornado Cash introduced anonymity mining. It was an incentive scheme to encourage more deposits in Tornado Cash pools, thereby increasing their anonymity sets. Tornado Cash rewarded participants a fixed amount of anonymity points (AP) based on how long they left their assets in a pool.

After withdrawing assets, users can claim Anonymity Points. The amount withdrawn is recorded in the transaction. If a user uses a single address to claim all of their anonymity points, one can calculate the exact number of Ethereum blocks that their assets were in the pool because the AP yields were public and fixed. If there is a unique deposit / withdrawal combination in a pool separated by this number of Ethereum blocks, the transactions are assumed linked. This is more likely when the deposit or withdrawal in the pair also claimed the AP. This heuristic is most effective if AP is being claimed for a single deposit and is harder to compute for multiple deposits. 

\subsection{Connection to Ethereum-based Clusters}

Unlike the Ethereum-wide heuristics (i.e., DAR and NODE), which find clusters of compromised \textit{addresses}, Tornado Cash heuristics find clusters of compromised \textit{transactions}. However, aside from address matching, a subset of these Tornado Cash heuristics can also be applied on the address level: given a cluster of compromised transactions, find the sender address for each transaction, and compute the unique set, removing any addresses in our list of known addresses. These clusters can be added to the ones discussed in Section~\ref{sec:eth}. This in turn, will further enrich the anonymity score.
\section{Analysis}
\label{sec:analysis}

We provide a brief quantitative analysis of the scale and efficacy of Tutela heuristics.
\begin{figure}[h!]
\includegraphics[width=\linewidth]{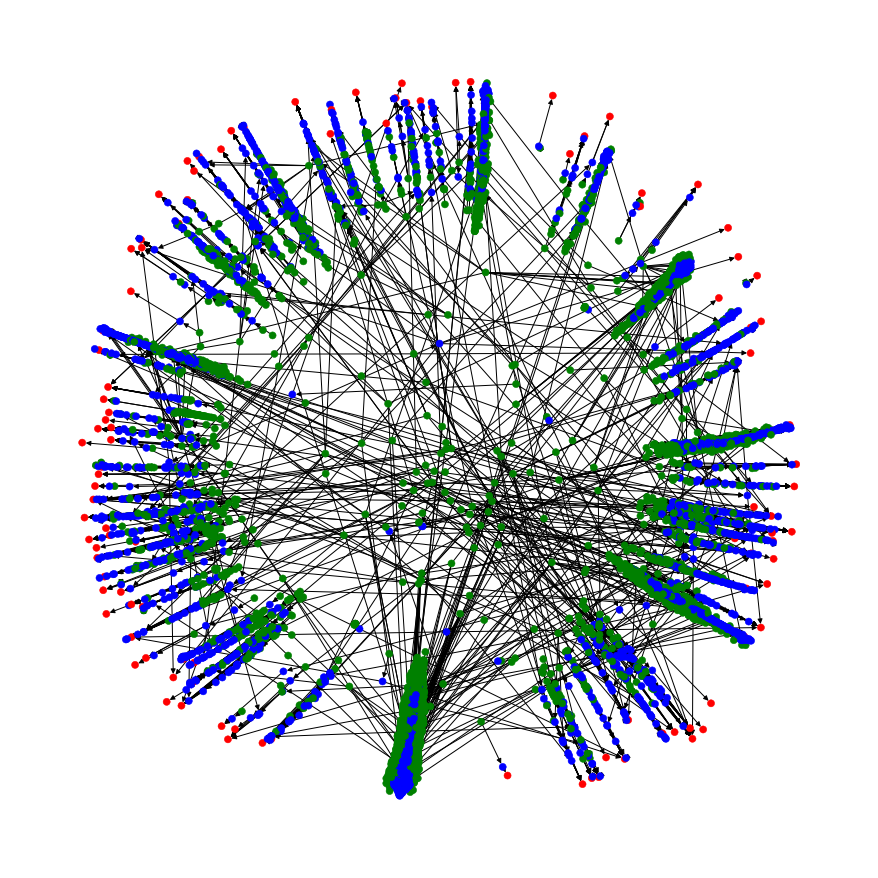}
\caption{10k subgraph of 26M graph created via deposit address reuse. Contains EOA (green), deposits (blue), and exchanges (red).}
\label{fig:dargraph}
\end{figure}
\subsection{Ethereum Heuristics}

Using DAR, we found 26M EOA addresses, resulting in 2.5M clusters of Ethereum addresses. The average cluster size contains 4.3 ($\pm$ 8.6) EOA addresses; the largest cluster contains 2.1k EOA addresses. Figure~\ref{fig:dargraph} shows a visualization of 10,000 random nodes from the DAR graph. In particular, we observe interesting structure with many small clusters scattered uniformly, balanced by several large clusters in the perimeter.

Next, we can measure the quality of the DAR clusters using a held out ``test set'' of known clustered addresses. We obtain such a set from \cite{beres2021blockchain} where 1,028 clusters of addresses (average size of 4.0 $\pm$ 3.6 EOA addresses per cluster) are derived from ENS names. We find a recall of 39.4\% using DAR. While there is room for improvement, we are cautiously optimistic that DAR is able to recover a nontrivial number without any knowledge of ENS names. This provides evidence for the generality of DAR clusters.

\begin{figure}[h!]
\includegraphics[width=\linewidth]{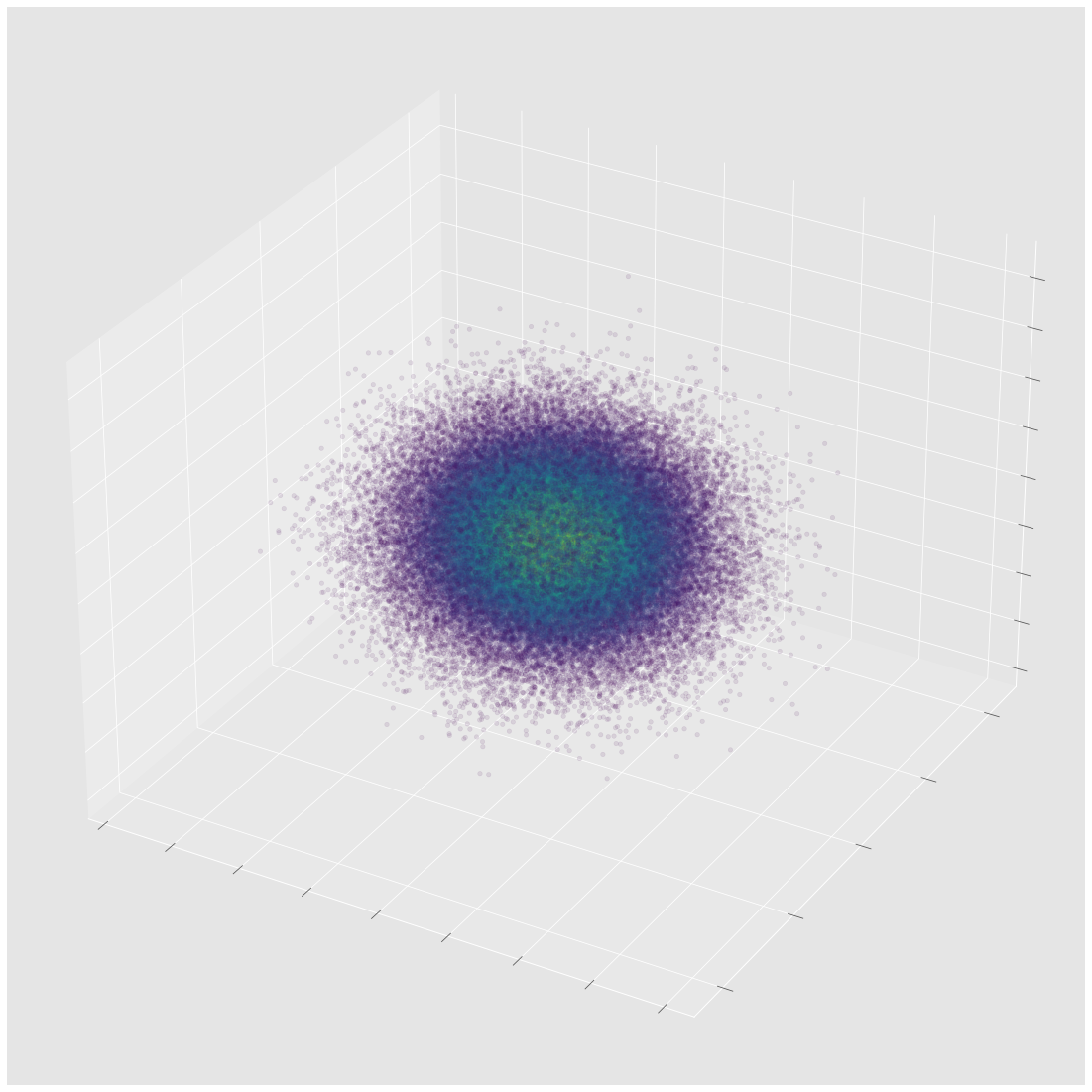}
\caption{100k random subset of 131M embedding of Ethereum addresses projected from to 3D using PCA.}
\label{fig:diff2vecgraph}
\end{figure}
\begin{figure}[h!]
\includegraphics[width=\linewidth]{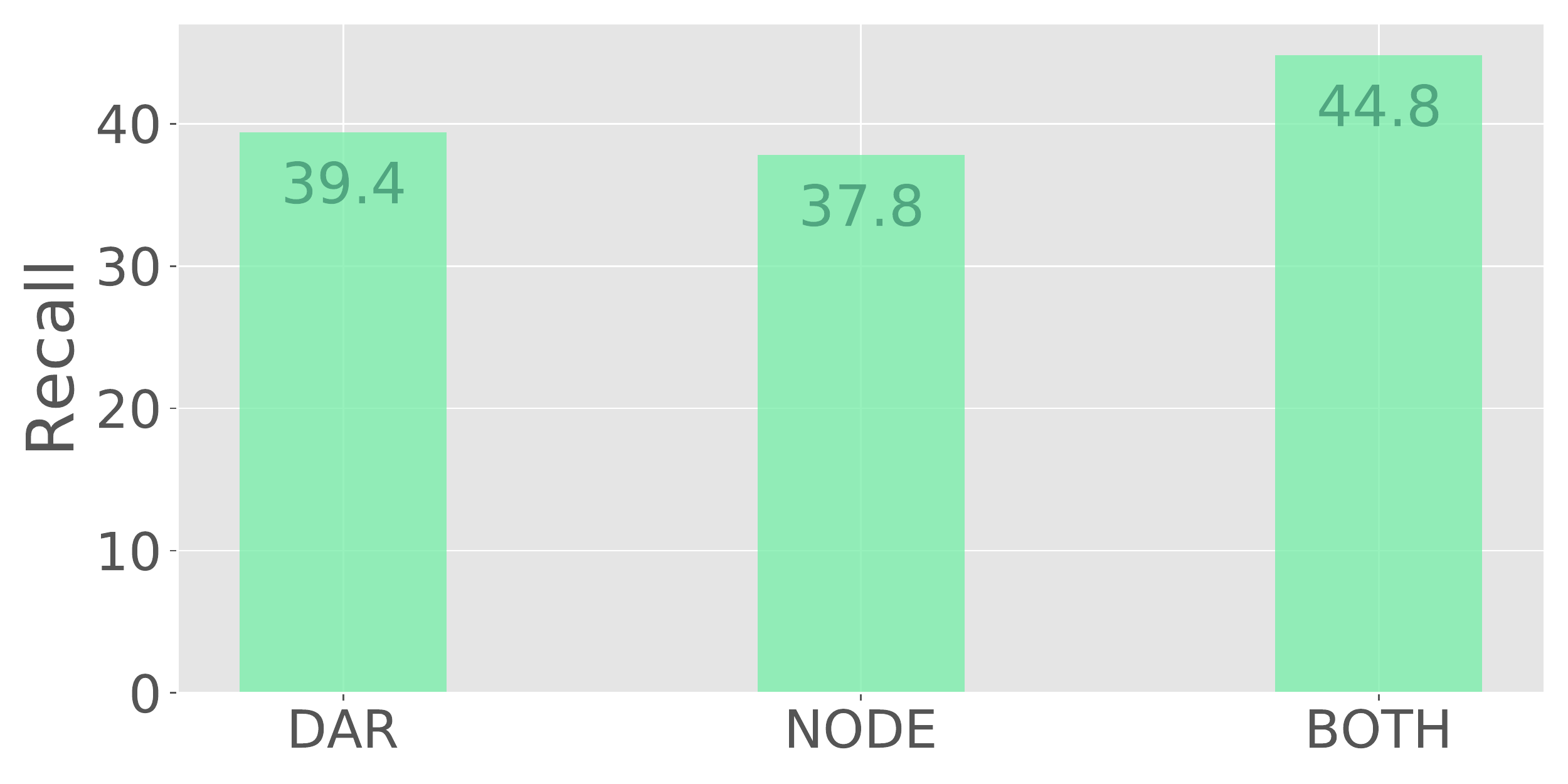}
\caption{Plot of the recall of held-out address clusters through ENS reveals using deposit address reuse (DAR), diff2vec (NODE), and the combination of both (BOTH). A higher recall represents a better heuristic.}
\label{fig:ensrecall}
\end{figure}

Using Node, we found 131M clusters of Ethereum addresses, with each cluster having exactly 9 members by design (10 including itself). Figure~\ref{fig:diff2vecgraph} shows a visualization of 100,000 random embeddings of Ethereum addresses from the NODE set, where embeddings are projected down to two dimensions using PCA (trained on a subset of 1M address embeddings). The color in Figure~\ref{fig:diff2vecgraph} shows the density, estimated using a Gaussian kernel, where a lighter (yellow) color represents higher density. That is, the figure shows a non-hollow ball of address embeddings. For any query address, Tutela returns the closest neighbors from the point cloud. Again, we access quality using the held-out known clusters, finding a recall of 37.8\% with NODE, two percent lower than DAR. However, as shown in Figure~\ref{fig:ensrecall}, when DAR and NODE are used in combination, the recall increases 7\% to 44.8\%, indicating that DAR and NODE find clusters of different ``types''. In Tutela, searching an address will return both clusters types.

\subsection{Tornado Cash Heuristics}

\begin{figure}[h!]
\includegraphics[width=\linewidth]{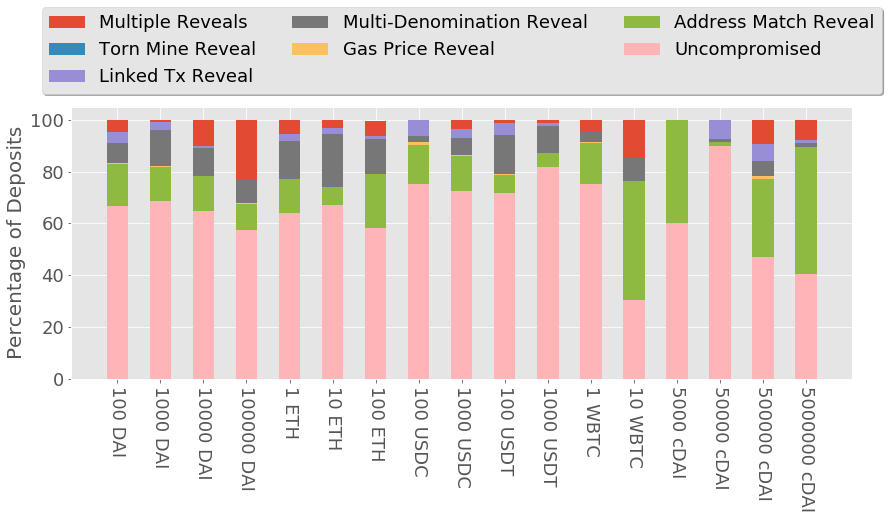}
\caption{Plot of the percentage of compromised versus uncompromised (pink) deposits by pool. }
\label{fig:tcashgraph}
\end{figure}

Of the 97.3k Tornado Cash equal user deposits, we found 42.8k are potentially compromised: 18.6K from the address match reveal, 102 from the unique gas price reveal, 18.9K from the linked ETH address reveal, 16.2K from the multi-denomination reveal, and 358 from the TORN mining reveal (with overlap between reveals). Splitting this by pool, we find the anonymity set to be reduced by 37\% ($\pm$ 15\%) on average. Figure~\ref{fig:tcashgraph} shows the uncompromised anonymity sets by pool.

We find that some of the pools could be heavily compromised (such as the cDAI and WBTC pools), whereas other pools are less effected (e.g. USDC). In summary, while many of the Tornado Cash heuristics are simple, they are quite powerful. These findings could help Tornado Cash developers and users alike, measure and understand the degree user privacy offered.

\section{Discussion and Future Work}
\label{sec:discussion}

We conclude with a few discussion points and final remarks. In particular, we present possible limitations and extensions to the proposed system.

\subsection{Limitations}

We acknowledge that heuristics are not perfect measures. In return for simplicity, there will likely be false positives in practice, e.g., addresses in a cluster that should not be there, or faithful Tornado Cash transactions labeled as compromised.

For the Ethereum heuristics, we emphasize that picking proper hyperparameters is a challenge. In DAR, the quality of the algorithm is very sensitive to the choice of maximum thresholds $\alpha$ and $\tau$. With too small thresholds, no clusters will be found; with too large thresholds, clusters will be low quality, containing many addresses they should not. Currently, the best practice is to tune these by hand. In NODE, the size of subgraph $l$ and the dimensionality $d$ similarly determine the resulting quality. The choice of $l$ should reflect the size of the full graph $G$: a bigger graph requires the embedding to summarize a larger neighborhood. Other NODE hyperparameters, such as window size or optimization choices, are less important to the final embedding. Running the DAR or NODE on Ethereum is computationally expensive, requiring both a large RAM and storage.

On the other hand, the Tornado Cash heuristics are much simpler than the Ethereum ones, and more deterministic.
However, given that only a small subset of Ethereum addresses are Tornado Cash users, they have limited applicability to the majority of potential Tutela users.


\subsection{Extensions}

As described, Tutela uses only on-chain data to access anonymity. However, extensions can be made to include off-chain data, such as from decentralized applications (e.g. DeFI, NFT, games, etc.), layer two data, external blockchains, and more. The inclusion of off-chain data could lead to more powerful de-anonymization attacks.

\subsection{Broader Impact}

There is a need for greater privacy on the blockchain to accelerate adoption. Consumers would not be willing to receive their salaries publically or have their online purchasing history in the public for all to see. Businesses would not pay suppliers on the blockchain if their competitors could see who and how much they pay for supplies. Similarly, investment funds want to keep their strategies private and not copied before their trades have even been recorded on-chain.

That said, blockchain privacy is a difficult issue to navigate. Currently, blockchain privacy and finality create opportunities for money laundering and nefarious activities, so privacy solutions combined with regulation will need to account for these considerations.
In the meantime, we hope that Tutela will help law-abiding blockchain users better protect themselves in the current ecosystem.

\section*{Acknowledgments}
We would like to acknowledge both the Tornado Cash team and the Convex Research team for their support. This work was funded by the Tornado Cash community bounty to develop anonymity tools to protect user privacy. We also thank the Tornado Cash community at large for their consistent feedback and patience through our application's development. Finally, we thank the Stanford Venture Studio for their support with compute credits.

\appendix
\section{Appendix}

\subsection{Tornado Cash Heuristic: Wallet Fingerprints}
We present a Tornado Cash heuristic for identifying compromised transactions that is not yet implemented in Tutela at the time of publication.

In the Ethereum ecosystem, users can select from numerous wallet softwares. Wallet fingerprints allow one to establish the wallet software a user sent their transaction from. Such wallet fingerprints could largely fragment the anonymity set of Tornado Cash users. For instance, a withdrawal transaction sent with wallet software $\mathcal{A}$ can only be indistinguishable among the deposit transactions sent by users of the same wallet software $\mathcal{A}$. 

Transaction fees (i.e., the $\mathsf{max\_fee}$ and $\mathsf{max\_priority\_fee}$ fields of a transaction) are typically set by the wallet software. These values are often computed by open-source algorithms that are different for each wallet. For instance, Blocknative's algorithm\footnote{See for more details: \url{https://www.blocknative.com/gas-estimator}.} uses the following: \[\mathsf{max\_fee}=\mathsf{base\_fee}+2\cdot\mathsf{max\_priority\_fee}\] Therefore, it is straightforward to assess the wallet software for many transactions. Assuming that accounts do not change their used wallet software, we can more confidently identify compromises.

\subsection{Improvements to Measuring Tornado Cash's True Anonymity Set Size}

The anonymity set size listed on Tornado Cash's webpage\footnote{https://tornadocash.eth.link} does not take into account compromised deposits. We can compute the uncompromised (or useful) anonymity set size as the total number of deposits in this pool removing revealed deposits from each of the heuristics above. This resulting number is a more faithful representation of the privacy provided by a Tornado Cash pool.

In this regard, the very definition of anonymity set offers a way of interpreting heuristics producing multiple deposit addresses as output for a single withdraw address input by measuring anonymity in terms of the Shannon entropy~\cite{diaz2002towards}. 

Concretely, we are able to quantify how much anonymity is lost after running a heuristic $\mathcal{H}$ for a given withdrawal address. Heuristics linking addresses by scanning ETH transactions outside the TC environment or linking addresses by matching deposit to withdraw portfolios typically produce a large number of candidate addresses: organizing these outputs as a non-empty set $\mathcal{C}$, they provide a natural "refined anonymity set" notion - of course, these outputs need not to include the actual deposit address, but in a worst-case scenario a clear interpretation can be given. Calling the anonymity set $\mathcal{D}$, $D$ its cardinality and recalling its definition, we can compute its Shannon entropy by simply taking the logarithm of its cardinality:

\begin{equation*}
H(\mathcal{D})=-\sum\limits_{d\in\mathcal{D}}{\frac{1}{D}\ln\Big(\frac{1}{D}\Big)}=\ln(D)
\end{equation*} 

This quantity is a central notion in classical information theory and represents how much "regularity" there is in a random variable: higher entropy will mean low information. In our context, the definition of anonymity set implies that any of its members is equally likely to be the deposit address actually linked to a given withdraw address and we interpret this fact as not having more information about the participant addresses of the actual transaction than the size of the TC anonymity set. A heuristic $\mathcal{H}$ (or combination of heuristics) producing a refined set of candidates $\mathcal{C}=\mathcal{C}(y)$ will then update our prior belief and potentially locate the actual deposit address in this set: now the conditional entropy $H(\mathcal{C})$ is subtracted from the a priori entropy resulting in a quantity known as conditional information gain:

\begin{equation*} 
\begin{aligned}
  I(X,X|Y=y)=H(X)-H(X|Y=y)=\\ =\ln(\lvert \mathcal{D}\rvert)-\ln(\lvert \mathcal{C}\rvert)
\end{aligned}
\end{equation*}

 This turns out to be equal to the Kullback-Leibler divergence between the prior and posterior probability mass functions for the anonymity sets. It measures how much information is gained upon running the heuristic. In this sense, larger sets $\mathcal{C}$ produce little information gain, whereas smaller ones produce larger information gains: in the case of $\mathcal{C}$ containing the actual deposit address corresponding to a given withdrawal, we interpret this information gain as a greater anonymity loss.

\bibliography{whitepaper}
\bibliographystyle{acl_natbib}

\end{document}